%% file: ms.tex
\documentclass[printer]{aa}
\usepackage{graphicx,natbib}
\usepackage{epic,eepic}

\usepackage{color}

\begin{document}
 
\input macros
\input aas_journals

\def\phxO{{\tt PHOENIX/1D}}
\def\phxT{{\tt PHOENIX/3D}}
\def\phx{{\tt PHOENIX}}

\title{A 3D radiative transfer framework: XIII. OpenCL implementation}

\titlerunning{3D radiative transfer framework XIII}
\authorrunning{Hauschildt and Baron}
\author{Peter H. Hauschildt\inst{1} and E.~Baron\inst{1,2,3}}

\institute{
Hamburger Sternwarte, Gojenbergsweg 112, 21029 Hamburg, Germany;
yeti@hs.uni-hamburg.de 
\and
Homer L.~Dodge Dept.~of Physics and Astronomy, University of
Oklahoma, 440 W.  Brooks, Rm 100, Norman, OK 73019 USA;
baron@ou.edu
\and
Computational Research Division, Lawrence Berkeley National Laboratory, MS
50F-1650, 1 Cyclotron Rd, Berkeley, CA 94720-8139 USA
}

\date{Received date \ Accepted date}

\abstract
{}
{We discuss an implementation of our 3D radiative
transfer (3DRT) framework with the OpenCL paradigm for
general GPU computing.
}
{We implement the kernel for solving the 3DRT problem 
in Cartesian coordinates with periodic boundary conditions
in the horizontal $(x,y)$ plane, including the construction
of the nearest neighbor $\Lstar$ and the operator splitting
step.
}
{We present the results of a small and a large test case 
and compare the timing of the 3DRT calculations for
serial CPUs and various GPUs.
}
{The latest available GPUs can lead to significant speedups
for both small and large grids compared to serial (single core)
computations.}


\maketitle

\section{Introduction}

In a series of papers \citet*[][hereafter: Papers I--VII]{3drt_paper1,
3drt_paper2, 3drt_paper3, 3drt_paper4, 3drt_paper5,3drt_paper6,3drt_paper7}, we
have described a framework for the solution of the radiative transfer equation
in 3D systems (3DRT), including a detailed treatment of scattering in continua
and lines with a non-local operator splitting method, and its use in the
general model atmosphere package \phx. 

The 3DRT framework discussed in the previous papers of this series requires
very substantial amounts of computing time due to the complexity of the
radiative transfer problem in strongly scattering environments. The standard
method to speed up these calculations is to implement parallel algorithms for
distributed memory machines using the MPI library
\citep{3drt_paper1,3drt_paper2}. The development of relatively
inexpensive graphic processing units (GPUs) with large numbers of
cores and the development of relatively easy to use programming models,
OpenCL and CUDA 
has made the use of GPUs attractive for the acceleration of scientific
simulations. 
GPUs are built to handle numerous lightweight parallel threads simultaneously
and offer theoretical peak performance far beyond that of current CPUs. However,
using them efficiently requires different programming methods and algorithms
than those employed on standard CPUs. We describe our first results of
implementing 
our 3DRT framework for a single geometry within the OpenCL \cite[]{OpenCL}
paradigm for generalized GPU and CPU computing.

\section{Method}

In the following discussion we use  notation of Papers I -- VII. The basic
framework and the methods used for the formal solution and the solution of the
scattering problem via  non-local operator splitting are discussed in detail in
these papers and will not be repeated here. Our implementation of the 3DRT
framework considers the most time consuming parts of the process --- the formal
solution and the solution of the operator splitting equations --- to obtain the
updated values of the mean intensities, whereas less time consuming parts of
the code (set-up, Ng acceleration, etc) are left to Fortran95 or C
implementations.  The OpenCL implementation of the 3DRT framework minimizes
data transfer between the host computer (CPU) and the GPU and keeps most of the
data locally on the GPU memory
Only the relevant input data (e.g., opacities) and the  results,
e.g., the mean intensities $J$ for all voxels, need to be transferred to and
from the GPU device.

\subsection{General purpose computing on graphic processors}

Using a GPU for numerical calculations requires special programming
environments.  While GPU manufacturers have provided programming 
environments for vendor-special hardware, e.g., CUDA \cite[]{CUDA} for
NVIDIA produced GPUs and {ATI Stream SDK \cite[]{STREAMS} (which has 
now been replaced by AMD APP \cite[]{AMDAPP} which
uses OpenCL) for AMD/ATI produced
GPUs.} The differences between these environments make programs
specific to them. Our applications need to
be extremely portable and thus having to code for multiple
vendor-specific environments is not 
acceptable. Fortunately, the open standard OpenCL \cite[]{OpenCL} was
designed to efficiently use not only GPUs but also modern multi-core
CPUs and other accelerators using a thread-centered programming
model. With OpenCL it is possible to run the same code on many different CPU
and GPU architectures. {There is a relatively minor cost of some
loss of performance when using OpenCL as compared to CUDA specific programming
models \cite[]{komatsu10}. This is insignificant for our application where portability is far more
important than the fraction of the theoretical peak performance attained for a
specific piece of hardware. At the present time, OpenCL is available for all
types of GPUs and CPUs, including accelerators such as the Cell Broadband
Engine (CBE), whereas CUDA is only available for NIVIA GPUs.  This is a major
concern for us as the technology is progressing rapidly and new hardware is
released frequently. Using a defined standard is, therefore, already
important to build a reliable code base that can easily be used in the
future. Maintaining several different codes for the same tasks in different
programming languages is, on the other hand, is costly in terms of human time
and also error prone. The disadvantage of this use of general standards is a
loss of performance. We consider this as a low price for the portability as
hardware features and performance increase dramatically with new hardware.
Therefore, we implemented our 3DRT framework in OpenCL for portability
reasons.}

The design of GPUs differs considerably from the design of CPUs, focusing
much more on simultaneous execution of many threads to hide memory access
latencies. In contrast to CPUs, branching is costly on GPUs and should
therefore be limited as much as possible. It is in many cases
faster to compute both branches of a decision and then select the correct
one afterwards rather than using conditional execution. This is not an
uncommon strategy and was used, for example, on the original CRAY
vector machines in the 1980s. In addition, 
GPUs provide better performance for regular memory access patterns.
The preferred programming
model for these GPU systems is a single-program, multiple-data (SPMD) type
scheme which is directly supported by OpenCL. Branching within a program
is allowed in this model, but often drastically reduces performance and thus
should be avoided.

\subsection{Implementation of the formal solution and $\Lstar$ computation}

As a first step, we have implemented the ``simplest'' formal solution kernel
in OpenCL. This is the kernel for Cartesian coordinates with 
periodic boundary conditions in the (horizontal) $x-y$ plane discussed
in \cite{3drt_paper3}. An OpenCL implementation of the formal solution
is in principle straight forward: For any given direction of photon 
propagation, all characteristics can be tracked simultaneously through 
the voxel grid, which corresponds in the OpenCL paradigm to a 2D kernel
(the characteristics are started on the inner or outer $x-y$ planes). 
The only substantial hurdle in the problem is that OpenCL (version 1.0 or
1.1) does not have facilities for atomic updates of floating point
variables. This is, however, necessary for the numerically correct
operation of a straight-forward implementation of the formal solution
for the calculation of the mean intensities and the $\Lstar$ operator.
Therefore, we have implemented a 2-pass kernel, where in the first
pass the intensities (etc) are computed and stored along the characteristics
which can be implemented with atomic operations on integer variables
when the results are stored per voxel rather than per characteristic.
In a second pass, these data are collected on a per voxel (3D) OpenCL
kernel. With this method, the results are identical (to the precision 
used in the OpenCL implementation) to the Fortran95 implementation. However,
the 2-pass method requires additional memory on the GPU to store the
intermediate 
results and the first pass generates complex memory access patterns which are
likely to limit performance on GPU based systems. 

\subsection{Implementation of the operator splitting step}

The second very time-consuming part of the 3DRT framework is the solution of
the operator splitting equations to compute the new values of the
mean intensities $J$ for all voxels. The Fortran95 code solves these
equations by Jordan or Gauss-Seidel iterations. In OpenCL, it is 
much simpler to implement the Jordan method as it requires less 
synchronization between threads than does the Gauss-Seidel method. The
OpenCL implementation uses a 3D kernel (all voxels simultaneously) and
locally buffers the $\Lstar$ during the iterations, which dramatically
speeds up the calculations.

\section{Results}

\subsection{Test case setup}

The test cases we have investigated follow the continuum tests used in
\cite{3drt_paper3}. In detail, we used the following configuration for
the tests: Periodic boundary conditions (PBCs) are realized in a plane
parallel slab. We use PBCs on the $x$ and $y$ axes, $z_{\rm max}$ is
at the outside boundary, $z_{\rm min}$ the inside boundary. The slab
has a finite optical depth in the $z$ axis.  The grey continuum
opacity is parameterized by a power law in the continuum optical depth
$\tstd$ in the $z$ axis. The basic model parameters are
\begin{enumerate}
\item The total thickness of the slab,  $z_{\rm max}- z_{\rm min} = 10^7\,$cm
\item The minimum optical depth in the continuum, $\t_{\rm std}^{\rm min} =
10^{-4}$ and the maximum optical depth in the continuum, $\t_{\rm std}^{\rm
max} = 10^8$.
\item An isothermal slab, $T=10^4$~K
\item  Boundary conditions: Outer boundary condition $I_{\rm bc}^{-}
  \equiv 0$ and  
inner boundary condition LTE diffusion for all wavelengths.
\item Parameterized coherent \& isotropic continuum scattering given by
\[
\chi_c = \epsilon_c \kappa_c + (1-\epsilon_c) \sigma_c
\]
with $0\le \epsilon_c \le 1$. 
$\kappa_c$ and $\sigma_c$ are the 
continuum absorption and scattering coefficients.
\end{enumerate}
For the tests presented here, we use $ \epsilon_c = 10^{-2}$
in order to allow single precision runs, for smaller epsilons
double precision is required for the solution of the operator
splitting equations.

 We have verified that the OpenCL calculations give
the same result as the Fortran95 CPU calculations for the 
formal solution (intensities), the mean intensities and 
the $\Lstar$ as well as that the solution of the 3D radiative
transfer problem is the same for both OpenCL and Fortran95. 
For OpenCL single precision calculations, the relative accuracy is 
about $10^{-5}$, which is acceptable for most calculations.

\subsection{Timing results}

\begin{figure*}
\centering
\resizebox{\hsize}{!}{\includegraphics[angle=00]{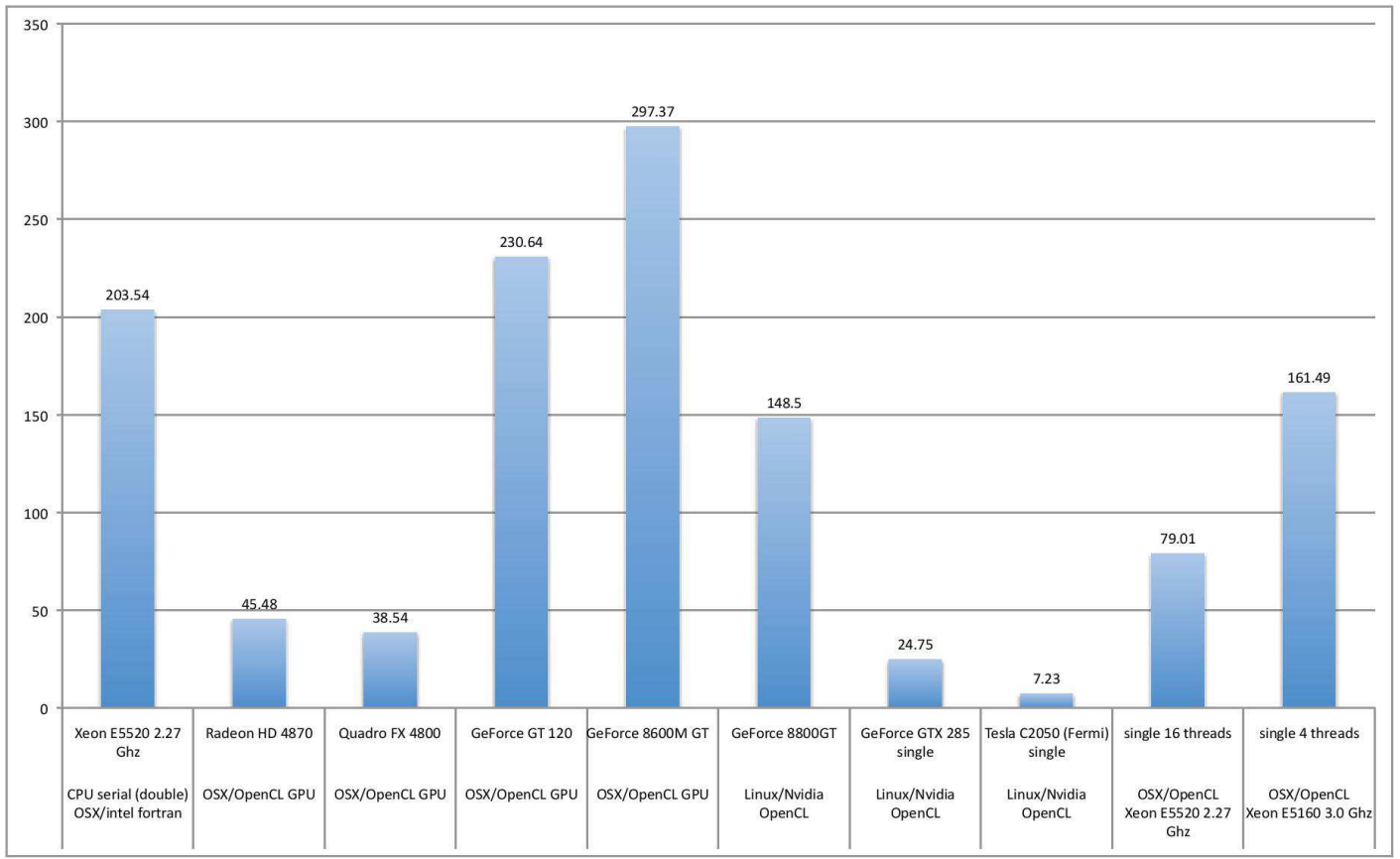}}
\caption{\label{fig:small:timing}
Timing of the small 3D radiative transfer test calculation on
CPUs (leftmost column), various GPUs with OpenCL, and multi-core CPUs using OpenCL.
The times are given in seconds of wallclock time.
}
\end{figure*}

\begin{figure*}
\centering
\resizebox{\hsize}{!}{\includegraphics[angle=00]{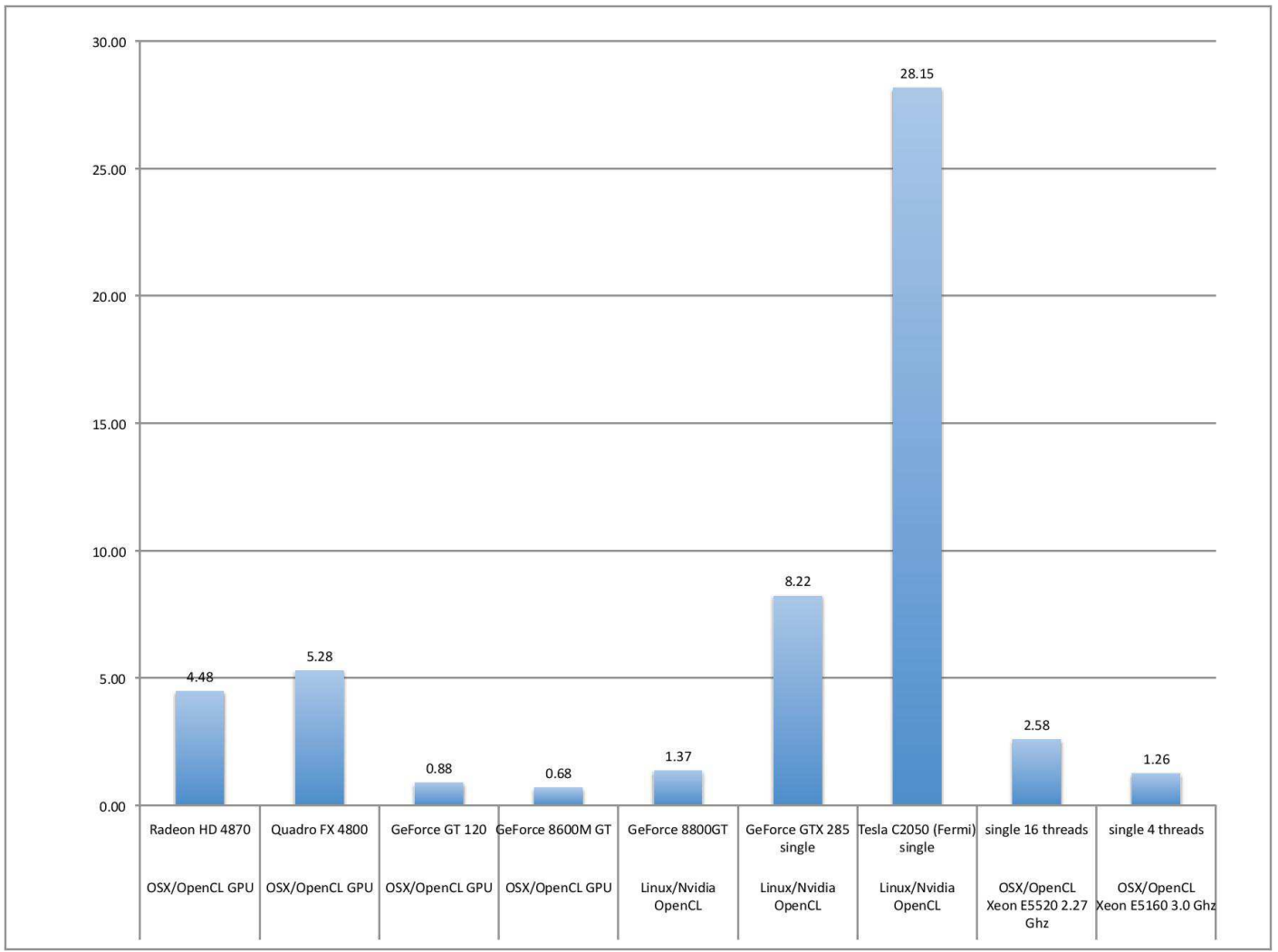}}
\caption{\label{fig:small:speedup}
Speed-ups of the small 3D radiative transfer test calculation for the OpenCL
implementation relative to the serial CPU run.
}
\end{figure*}

\begin{figure*}
\centering
\resizebox{\hsize}{!}{\includegraphics[angle=00]{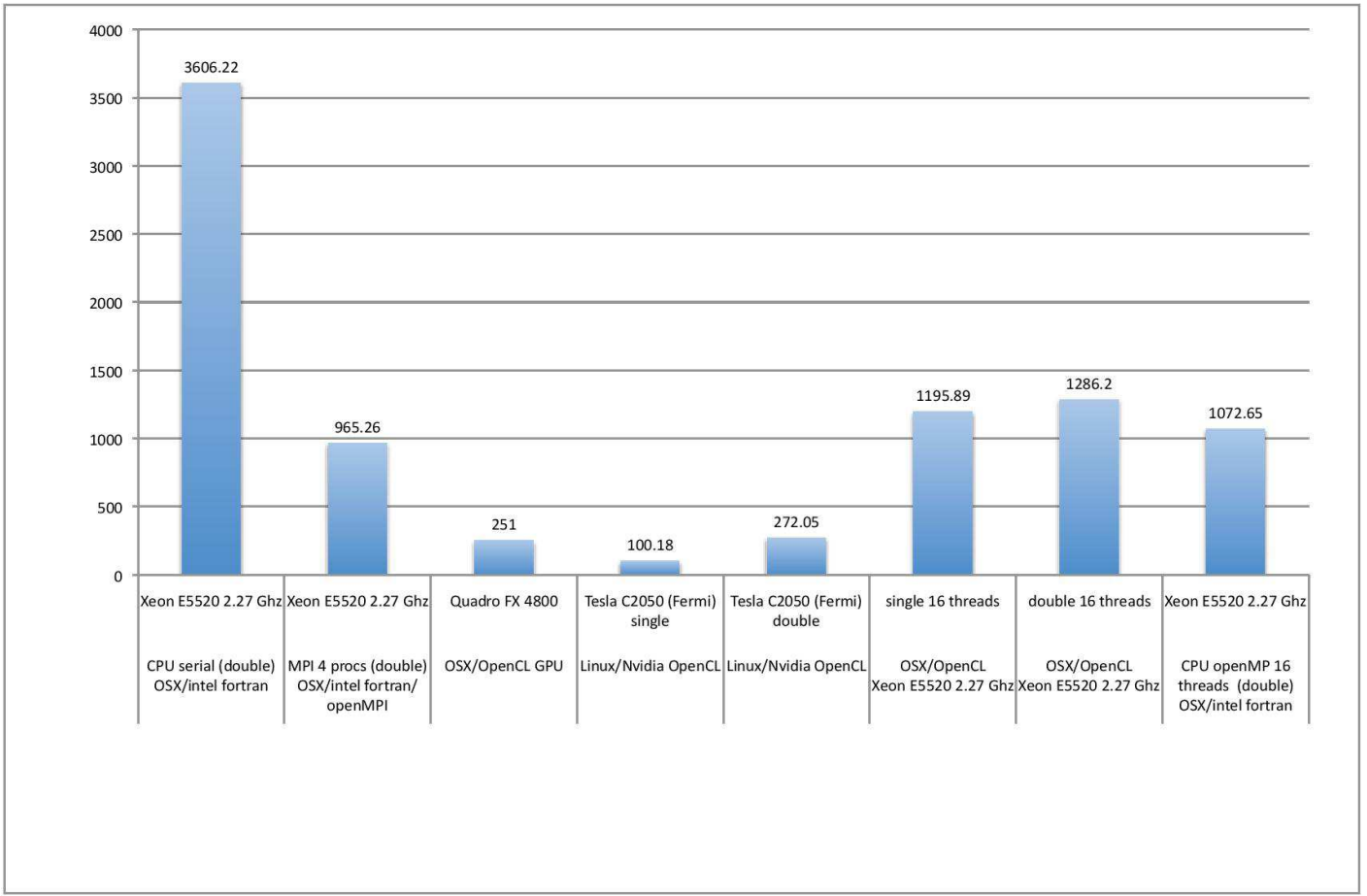}}
\caption{\label{fig:large:timing}
Timing of the large 3D radiative transfer test calculation on CPUs (leftmost
column), various GPUs with OpenCL, and multi-core CPUs using OpenCL, MPI and
OpenMP. The MPI calculation was run on 4 cores (1 CPU), the OpenMP run used 16
threads (8 cores, incl. hyperthreading) to be comparable to the OpenCL CPU run.
The times are given in seconds of wallclock time.
}
\end{figure*}

\begin{figure*}
\centering
\resizebox{\hsize}{!}{\includegraphics[angle=00]{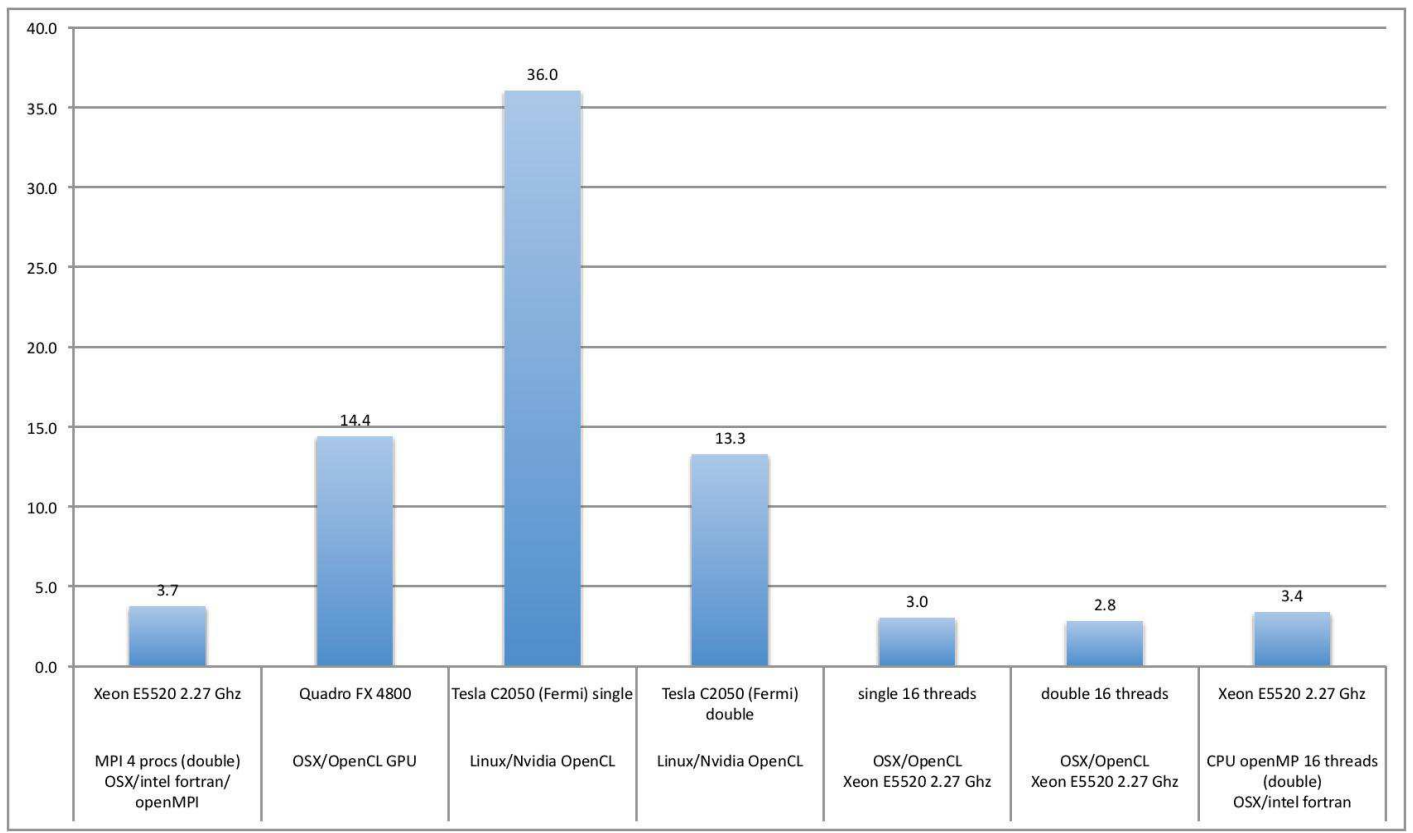}}
\caption{\label{fig:large:speedup}
Speed-ups of the large 3D radiative transfer test calculation for the OpenCL,
OpenMP, and MPI implementations relative to the serial CPU run. The MPI
calculation was run on 4 cores (1 CPU), the OpenMP run used 16 threads (8 cores,
incl. hyperthreading) to be comparable to the OpenCL CPU run.
}
\end{figure*}

In Figures~\ref{fig:small:timing}--\ref{fig:large:speedup}
we show the timing and speed-up results for small and large
test cases. The difference between the tests is simply the 
size of the voxel grid. In the small case, we use
$65^3 = 274,625$ voxels, whereas the large test case uses a grid
with $129^2*193 = 3,211,713$ voxels. The small test
cases uses 95.3\,MB OpenCL memory in single precision 
(165.5\,MB double precision) and thus fits easily in 
GPU devices with little memory, whereas the large
test case uses 1.1\,GB OpenCL memory in single
precision (1.9\,GB in double precision) and can thus only
be used on high-end GPU devices. The tests were 
run on a variety of systems, from laptops with 
low-end GPUs to Xeon-based systems with dedicated
GPU based numerical accelerator boards. For the comparisons
in the figures, we have selected the fastest CPU run 
as the serial baseline for all comparisons. The systems used in 
the tests are:
\begin{enumerate}
\item serial CPU: Mac Pro with OSX 10.6.4 and Intel Fortran 11.1, the
  CPU is a Xeon E5520 with 2.27\,GHz clock-speed
\item MPI: 4 processes on Mac Pro with OSX 10.6.4 and Intel Fortran
  11.1, the CPU is a Xeon E5520 (4 cores) with 2.27\,GHz clock-speed
\item AMD/ATI Radeon HD 4870 GPU with 512\,MB RAM on a Mac Pro with
  OSX 10.6.4 OpenCL
\item NVIDIA GeForce 8600M GT GPU with 512\,MB RAM on a MacBook Pro
  laptop with OSX 10.6.4 OpenCL
\item NVIDIA GeForce GT 120 GPU with 512\,MB RAM on a Mac Pro with OSX
  10.6.4 OpenCL
\item NVIDIA Quadro FX 4800 GPU with 1536\,MB RAM on a Mac Pro with
  OSX 10.6.4 OpenCL
\item NVIDIA GeForce 8800GT GPU with 512\,MB RAM on a Linux PC with
  NVIDIA OpenCL (OpenCL 1.1, CUDA 3.2.1)
\item NVIDIA GeForce GTX 285 GPU with 1024\,MB RAM on a Linux PC with
  NVIDIA OpenCL (OpenCL 1.0, CUDA 3.0.1)
\item NVIDIA Tesla C2050 Fermi-GPU with 2687\,MB RAM on a Linux PC
  with NVIDIA OpenCL (OpenCL 1.1, CUDA 3.2.1)
\item Intel Xeon E5520 at 2.27\,GHz clock-speed on a Mac Pro with OSX
  10.6.4, OpenCL (16 OpenCL threads)
\item Intel Xeon E5520 at 2.27\,GHz clock-speed on a Mac Pro with OSX
  10.6.4, Intel Fortran 11.1 OpenMP (16 OpenMP threads)
\end{enumerate}
All these system were used for the small test case, the large test 
case could only be run on the serial CPU, the 2 OpenCL CPU runs, and
on the Quadro FX 4800 and Tesla C2050 GPUs due to memory constraints.

The results for the small test case show that low-end GPUs (GeForce GT 120,
GeForce 8xxx) do not provide significant speed-up compared to serial CPU
calculations. However, compared to a laptop class CPU they can be useful
(the GeForce 8600M GT reaches about the speed of the serial CPU of the laptop,
an Intel T9300 CPU at 2.5GHz) as they can be used in parallel with the CPU
(e.g., to offload formal solutions for visualizations from the CPU). Medium
grade GPUs (Radeon HD 4870 or Quadro FX 4800) give speed-ups of the order of
4-5 compared to CPUs, which is already quite useful for small scale
calculations on workstations.  High-end GPUs (GeForce GTX 285, Tesla C2050)
deliver substantially larger speed-ups for the small test case, a
factor of 28 for the Fermi-GPU based accelerator is very significant for
calculations.

For the large case, which is close to the size of a real production
calculation, we show the timing results in Figs.~\ref{fig:large:timing} and
\ref{fig:large:speedup}. The memory requirements of the calculations now limit
the tests to the CPUs (serial and OpenCL) and the Quadro FX 4800 and Tesla
C2050 devices. For the OpenCL runs on CPUs and the Tesla C2050 runs we also
include the results for double precision OpenCL calculations. In this test, the
GPUs deliver larger speed-ups, up to a factor of 36 for the Tesla C2050 device.
Using double precision reduces the speed-up to about a factor of 13 (a factor
of about 2.7 slower than single precision), which is presumably due to more
complex memory accesses and less efficient double precision hardware on the
Fermi GPU.  Running OpenCL on CPUs is still not efficient compared to running
MPI code, but the  timings are essentially  the same regardless of single or
double precision. OpenCL is about as efficient as using OpenMP shared
memory parallelization with the same number of threads. Therefore, 
OpenCL can be used as a more versatile replacement for OpenMP code.
{It has been suggested that ultimately GPUs may be only a factor of 2-3
faster than multi-core CPUs \citep{lee10}, therefore, a careful split 
of the computational work between CPUs and GPUs is probably the most 
efficient way to use these systems.}

Comparing our results to others in the literature is not easy, since most other
physics and astronomy applications solve problems that have very different
computational characteristics 
{with different levels of difficulty in their parallelization
on SMP machines, e.g. N-body problems \cite[]{capuzzo11,GBPZ10}
where speedups can a factor of 100 or more when a clever strategy is adopted for the evaluation of
the pairwise force in the system by direct summation.
On the other hand, }
in molecular dynamics problems speedups of 20--60 are considered acceptable
\citep{multi_core09}. The extremely non-local nature of the radiative transfer
problem makes the kernels extremely complex. Therefore, in order to retain
numerical accuracy and portability, some fraction of the theoretical speedup
must be sacrificed.

We note that we have chosen OpenCL for its portability and we have
so far not tried to fully optimize our kernels for the specific architecture as new features
and fundamental changes are introduced very frequently into new hardware {
and better OpenCL compilers will reduce the importance of hand-tuning 
the OpenCL code reported by \cite{komatsu10}}.
Studies have shown that hand-tuned optimizations can lead to OpenCL performance
approaching that given by using vendor specific software
\citep{weber2011,komatsu10}, but in this still early state of general computing
on GPUs this will change with new versions of the OpenCL framework and better
OpenCL compilers. {One of the main issues concerning further
optimization is an inherent problem of the formal solution: for each solid angle
(direction) any characteristic passes through a large fraction of the voxel grid, resulting
in highly complex memory access patterns that also vary from one direction
to another and from one characteristic to another. This is a basic 
feature of radiative transfer. Using approaches that maximize data locality work on
CPUs (with a small number of cores) but are very inefficient on GPUs as 
many PEs may be idle at any given time (again, depending on the direction).
We have done a number of experiments with different approaches and found
that the parallel tracking implementation described above is a good overall 
compromise that keeps many OpenCL threads active and allows the GPU hardware
to mask many memory latencies. On CPUs, the impact is even smaller
as the number of cores tends to be small and complex memory access patterns 
are handled more efficiently than for a single thread on a GPU. With this the 
algorithm performs best on newer GPU hardware compared to older hardware,
which is encouraging for future devices. A promising venue for further optimization
requires the availability of atomic updates for floating point numbers 
in OpenCL, this could remove the need for a two pass approach in the formal
solution and may improve performance.}

\section{Summary and Conclusions}

We have described first results we have obtained implementing our 3DRT
framework in OpenCL for use on GPUs and similar accelerators. The
results for 3D radiative transfer in Cartesian coordinates with
periodic boundary conditions show that high-end GPUs can results in
quite large speed-ups compared to serial CPUs and are thus useful to
accelerate complex calculations.  This is in particular useful for
clusters where each node has one GPU device and where calculations can
be domain-decompositioned with a one node granularity. Large scale
calculations that require a domain-decomposition larger than one node
are more efficient on large scale supercomputers with 1000's of cores
as data transfer required for simultaneous use of multiple GPUs on
multiple nodes will dramatically reduce performance.  Even medium-end
or low-end GPUs can be useful to offload calculations from the CPU to
speed up the overall calculations.

\begin{acknowledgements}
  This work was supported in part by DFG GrK 1351 and SFB 676, as well
  as NSF grant AST-0707704, US DOE Grant DE-FG02-07ER41517 and NASA
  Grant HST-GO-12298.05-A.  Support for Program number
  HST-GO-12298.05-A was provided by NASA through a grant from the
  Space Telescope Science Institute, which is operated by the
  Association of Universities for Research in Astronomy, Incorporated,
  under NASA contract NAS5-26555.  The calculations presented here
  were performed at the H\"ochstleistungs Rechenzentrum Nord (HLRN)
  and at the National Energy Research Supercomputer Center (NERSC),
  which is supported by the Office of Science of the U.S.  Department
  of Energy under Contract No. DE-AC03-76SF00098.  We thank all these
  institutions for a generous allocation of computer time.
\end{acknowledgements}

\bibliography{yeti,radtran,rte_paper2,stars,OpenCL,paper9}

\clearpage

\end{document}

%% file: macros.tex
%
%
\def\valid{}    

\font\caps=cmcsc10                  
\font\dunh=cmdunh10  at 12.0 true pt 
\font\dunhs=cmdunh10 
\font\vbold=cmbx10 scaled \magstep1 
\font\sevenbf=cmbx7
\font\sevenit=cmti7
\font\Kapi=cmr17

\def\MEV{DOME}
\def\RTE{equation of radiative transfer}
\def\etal{{et al}}
\def\HW{H\&W}
\def\OK{O\&K}
\def\ok{O\&K}
\def\RH{R\&H}

\def\ibmrs{\hbox{\tt RS/6000}}
\def\hp{\hbox{\tt HP~9000}}
\def\dec{\hbox{\tt DEC~5000}}
\def\axp{\hbox{\tt AXP}}
\def\ibmmf{\hbox{\tt IBM~3090}}
\def\ibmpc{\hbox{\tt 486DX}}
\def\cray{\hbox{\tt Cray 2}}
\def\ymp{\hbox{\tt YMP}}
\def\nec{\hbox{\tt NEC}}

\def\g{\gamma}
\def\b{\beta}
\def\m{\mu}
\def\e{\epsilon}
\def\n{\nu}
\def\l{\lambda}
\def\L{\Lambda}
\def\K{{\rm K}}
\def\logg{{\log(g)}}
\def\t{\tau}
\def\pder#1#2{{\partial #1 \over \partial #2}}
\def\div#1#2{{#1\over #2}}
\def\rout{\ifmmode{r_{\rm out}}\else\hbox{$r_{\rm out}$}\fi}
\def\tmax{\ifmmode{\tau_{\rm max}}\else\hbox{$\tau_{\rm max}$}\fi}
\def\tstd{\ifmmode{\tau_{\rm std}}\else\hbox{$\tau_{\rm std}$}\fi}
\def\vmax{\ifmmode{v_{\rm max}}\else\hbox{$v_{\rm max}$}\fi}
\def\muE{\ifmmode{\mu_{\rm E}}\else\hbox{$\mu_{\rm E}$}\fi} 
\def\pE{\ifmmode{p_{\rm E}}\else\hbox{$p_{\rm E}$}\fi} 
\def\bmax{\ifmmode{\b_{\rm max}}\else\hbox{$\b_{\rm max}$}\fi}
\def\kms{\hbox{$\,$km$\,$s$^{-1}$}}
\def\ergs{\hbox{$\,$erg$\,$s$^{-1}$}}
\def\kpc{\hbox{$\,$kpc} }
\def\ang{\hbox{\AA}}
\def\Msun{\hbox{$\,$M$_\odot$} }
\def\Lsun{\hbox{$\,$L$_\odot$} }
\def\Teff{\hbox{$\,T_{\rm eff}$} }
\def\alog#1{\times 10^{#1}}
\def\rin{\hbox{$r_{\rm in}$} }
\def\rout{\hbox{$r_{\rm out}$} }

\def\lstar{\ifmmode{\Lambda^*}\else\hbox{$\Lambda^*$}\fi} 
\def\Lstar{\ifmmode{\Lambda^*}\else\hbox{$\Lambda^*$}\fi} 
\def\Rop{\ifmmode{[R_{ij}]}\else\hbox{$[R_{ij}]$}\fi}
\def\Rij{\Rop}
\def\Rji{\ifmmode{[R_{ji}]}\else\hbox{$[R_{ji}]$}\fi}
\def\Rstar{\ifmmode{[R_{ij}^*]}\else\hbox{$[R_{ij}^*]$}\fi}
\def\Rijstar{\Rstar}
\def\Rjistar{\ifmmode{[R_{ji}^*]}\else\hbox{$[R_{ji}^*]$}\fi}
\def\DRji{\ifmmode{[\Delta R_{ji}]}\else\hbox{$[\Delta R_{ji}]$}\fi}
\def\DRij{\ifmmode{[\Delta R_{ij}]}\else\hbox{$[\Delta R_{ij}]$}\fi}

\def\Jb{{\bar J}}
\def\Jbar{{\bar J}}
\def\Jnew{{\bar J_{\rm new}}}
\def\Jold{{\bar J_{\rm old}}}
\def\Jfs{{\bar J_{\rm fs}}}
\def\Snew{{S_{\rm new}}}
\def\Sold{{S_{\rm old}}}
\def\Amat{\mat{A}}             

\def\ns{\ifmmode{N_{\rm s}}          
        \else\hbox{$N_{\rm s}$}\fi}
\def\ion#1{\hbox{ #1}}         

\def\peq{\mathbin{\hbox{$+$}\hbox{$=$}}}

\def\mat#1{{\bf #1}}     
\def\vek#1{{#1}}         

\newcount\eqcount
\eqcount=0
\def
  \nummer{
    \global\advance\eqcount by 1
    (\the\eqcount)
  }

\def
  \numadv{
    \global\advance\eqcount by 1
  }

\def
   \numout#1{
     (\the\eqcount #1)
  }

\def\ivek#1#2{\ifmmode{\vek{I}^{#1}_{#2}}
        \else\hbox{$\vek{I}^{#1}_{#2}$}\fi}

\def\ip#1{\ivek{+}{#1}}      
\def\im#1{\ivek{-}{#1}}      

\def\tmat#1#2{\ifmmode{\mat{t}^{#1}_{#2}}
        \else\hbox{$\mat{t}^{#1}_{#2}$}\fi}
\def\rmat#1#2{\ifmmode{\mat{r}^{#1}_{#2}}
        \else\hbox{$\mat{r}^{#1}_{#2}$}\fi}
\def\bvek#1#2{\ifmmode{\beta^{#1}_{#2}}
        \else\hbox{$\beta^{#1}_{#2}$}\fi}

\def\tpi#1{\tmat{+}{#1}}
\def\tmi#1{\tmat{-}{#1}}
\def\rmi#1{\rmat{-}{#1}}
\def\rpi#1{\rmat{+}{#1}}
\def\bpi#1{\bvek{+}{#1}}
\def\bmi#1{\bvek{-}{#1}}

\def\tp{\tmat{+}{}}          
\def\tm{\tmat{-}{}}          
\def\rmm{\rmat{-}{}}         
\def\rp{\rmat{+}{}}          
\def\bp{\bvek{+}{}}          
\def\bm{\bvek{-}{}}          
\def\tpm{\tmat{\pm}{}}       
\def\rpm{\rmat{\pm}{}}       
\def\bpm{\bvek{\pm}{}}       

\def\lp{\ifmmode{\lambda^+_\tau}           
        \else\hbox{$\lambda^+_\tau$}\fi}
\def\lm{\ifmmode\lambda^-_\tau             
        \else\hbox{$\lambda^-_\tau$}\fi}

%% file: aas_journals.tex
%
%
%
%



\def\aasref@jnl#1{{\rm #1}}

\def\aj{\aasref@jnl{AJ}}                   
\def\araa{\aasref@jnl{ARA\&A}}             
\def\apj{\aasref@jnl{ApJ}}                 
\def\apjl{\aasref@jnl{ApJ}}                
\def\apjs{\aasref@jnl{ApJS}}               
\def\ao{\aasref@jnl{Appl.~Opt.}}           
\def\apss{\aasref@jnl{Ap\&SS}}             
\def\aap{\aasref@jnl{A\&A}}                
\def\aapr{\aasref@jnl{A\&A~Rev.}}          
\def\aaps{\aasref@jnl{A\&AS}}              
\def\azh{\aasref@jnl{AZh}}                 
\def\baas{\aasref@jnl{BAAS}}               
\def\jrasc{\aasref@jnl{JRASC}}             
\def\memras{\aasref@jnl{MmRAS}}            
\def\mnras{\aasref@jnl{MNRAS}}             
\def\pra{\aasref@jnl{Phys.~Rev.~A}}        
\def\prb{\aasref@jnl{Phys.~Rev.~B}}        
\def\prc{\aasref@jnl{Phys.~Rev.~C}}        
\def\prd{\aasref@jnl{Phys.~Rev.~D}}        
\def\pre{\aasref@jnl{Phys.~Rev.~E}}        
\def\prl{\aasref@jnl{Phys.~Rev.~Lett.}}    
\def\pasp{\aasref@jnl{PASP}}               
\def\pasj{\aasref@jnl{PASJ}}               
\def\qjras{\aasref@jnl{QJRAS}}             
\def\skytel{\aasref@jnl{S\&T}}             
\def\solphys{\aasref@jnl{Sol.~Phys.}}      
\def\sovast{\aasref@jnl{Soviet~Ast.}}      
\def\ssr{\aasref@jnl{Space~Sci.~Rev.}}     
\def\zap{\aasref@jnl{ZAp}}                 
\def\nat{\aasref@jnl{Nature}}              
\def\iaucirc{\aasref@jnl{IAU~Circ.}}       
\def\aplett{\aasref@jnl{Astrophys.~Lett.}} 
\def\apspr{\aasref@jnl{Astrophys.~Space~Phys.~Res.}}
\def\bain{\aasref@jnl{Bull.~Astron.~Inst.~Netherlands}} 
\def\fcp{\aasref@jnl{Fund.~Cosmic~Phys.}}  
\def\gca{\aasref@jnl{Geochim.~Cosmochim.~Acta}}   
\def\grl{\aasref@jnl{Geophys.~Res.~Lett.}} 
\def\jcp{\aasref@jnl{J.~Chem.~Phys.}}      
\def\jgr{\aasref@jnl{J.~Geophys.~Res.}}    
\def\jqsrt{\aasref@jnl{J.~Quant.~Spec.~Radiat.~Transf.}}
\def\memsai{\aasref@jnl{Mem.~Soc.~Astron.~Italiana}}
\def\nphysa{\aasref@jnl{Nucl.~Phys.~A}}   
\def\physrep{\aasref@jnl{Phys.~Rep.}}   
\def\physscr{\aasref@jnl{Phys.~Scr}}   
\def\planss{\aasref@jnl{Planet.~Space~Sci.}}   
\def\procspie{\aasref@jnl{Proc.~SPIE}}   

\let\astap=\aap
\let\apjlett=\apjl
\let\apjsupp=\apjs
\let\applopt=\ao